\documentclass[12pt,preprint]{aastex}

\shorttitle{SN 1998bw}
\shortauthors{Clocchiatti et al.}

\slugcomment{Submitted to, and accepted by, Astronomical Journal}

\begin{document}

\title{The Ultimate Light Curve of SN 1998bw/GRB~980425}

\author{
Alejandro Clocchiatti,\altaffilmark{1}
Nicholas B. Suntzeff,\altaffilmark{2,5}
Ricardo Covarrubias,\altaffilmark{3,5}
and
Pablo Candia,\altaffilmark{4,5}
}

\altaffiltext{1}{Pontificia Universidad Cat\'{o}lica de Chile,
Departamento de Astronom\'{\i}a y Astrof\'{\i}sica,
Casilla 306, Santiago 22, Chile; {aclocchi@astro.puc.cl}}
\altaffiltext{2}{Department of Physics \& Astronomy, Mitchell Institute for Fundamental Physics \& Astronony, Texas A\&M University,
College Station, TX 77843; {nsuntzeff@tamu.edu}}
\altaffiltext{3}{Australian Astronomical Observatory, P.O. Box 296, Epping, NSW 1710, Australia; {ricardo@aao.gov.au}}
\altaffiltext{4}{Gemini Observatory, Casilla 603, La Serena, Chile;
{pcandia@gemini.edu}}
\altaffiltext{5}{Staff astronomers, Cerro Tololo Inter-American Observatory, National
Optical Astronomy Observatory, which are operated by the Association of
Universities for Research in Astronomy, under contract with the National
Science Foundation.}

\begin{abstract}
We present multicolor light curves of SN 1998bw which appeared
in ESO184-G82 in close temporal and spacial association with GRB~980425.
They are based on observations done at Cerro Tololo
Inter-American Observatory and data from the literature.
The CTIO photometry reaches $\sim$86 days after the GRB in $U$ and $\sim$160
days after the GRB in $BV(RI)_C$.
The observations in $U$ extend by
about 30 days the previously known coverage,
and determine the slope of the early exponential tail.

We calibrate a large set of local standards in common with those
of previous studies and use them to transform published
observations of the SN to our realization of the standard photometric system.
We show that the photometry from different sources merges smoothly and
provide a unified set of 300 observations of the SN in five bands.
Using the extensive set of spectra in public domain we compute
extinction and $K$ corrections, and build quasi-bolometric unreddened
rest frame light curves.
We provide low degree piecewise spline fits to these light curves
with daily sampling.
They reach
$\sim$86 rest frame days after the GRB with $U$ band coverage, and 
$\sim$498 rest frame days after the GRB without $U$.
\end{abstract}

\keywords{supernovae: general --- GRB: general ---
supernovae: individual(\objectname{SN 1998bw})
GRB: individual(\objectname{GRB 980425})
}

\section{Introduction}

Stripped envelope core--collapse Supernovae (SNe) remain one of the open frontiers for basic observational work on SNe.
Without the appeal of Type Ia and Type~II SNe which, through different
techniques can be used as distance estimators, their follow up is typically
neglected when they are discovered in SN searches, unless they are particularly
bright or display puzzling peculiarities.
As a consequence, progress in this area comes at a slower pace.

After a peak in activity during 1993--1995, which resulted mainly
from the impact of the bright and well observed SNe~1993J and 1994I, the field of these ``exotic''
SNe entered a time of slower productivity.
In 1998, however, a long-duration $\gamma$--ray burst (GRB) provided a new insight
into core--collapse SNe, and gave new incentives to justify their follow up.
The association of long-duration, soft-spectrum  GRBs with SNe has been explored
in many publications \citep{wandb06, hartmann10, wanda10}.
It is now considered that most
long-duration soft-spectrum GRBs are accompanied by massive stellar explosions.

One of the events that made a significant contribution to establish
the connection GRB-SNe was SN~1998bw / GRB980425 \citep{getal98}.
It was the target of extensive coverage both in terms of wavelength range
and time span, and subject of many theoretical studies to understand the
nature of the progenitor and details of the certainly asymmetric
explosion.

SN~1998bw was discovered after a search triggered by GRB980425, 1.6 arc-minutes
away from the center of the 8 arc-minute error box of the Wide Field Camera of
BeppoSAX \citep{getal98}.
After some initial confusion, it was recognized as a peculiar Type~Ib/c SNe
\citep{setal98}.
It appeared superimposed on a complicated background, as became clear with the
exquisite resolution of {\em HST} \citep{petal01}.
Hence, while simple aperture photometry, or Point Spread
Function (PSF) fitting photometry, were reasonable approaches when the SN was bright,
more sophisticated techniques became necessary, when it was fading on.
\citet{getal98} presented the first set of observations in the
optical and near infrared passbands.
Their light curves start as early as $\sim$17 hours after the GRB, in $V$ and $R_C$, and
extend up to $\sim$57 days after the GRB with $UBV(RI)_C$ coverage. They use simple aperture photometry
to estimate the SN brightness.
\citet{mands99} present $BVI_C$ photometry, from $\sim$64 up to $\sim$187 days after the GRB.
\citet{petal01} present an extensive set of observations, including spectroscopy, spectropolarimetry, and photometry.
The later spans the range from 323 up to 426 days after the GRB, and the brightness is measured with Point Spread
Function (PSF) fitting photometry using a software specifically designed for analyzing point sources superimposed
on bright, spatially variable, backgrounds.
\citet{soetal02} present $BV(RI)_C$ photometry that extends from 140 up to more than 500 days after the GRB, and
includes data from a reanalysis of the images used by \citet{petal01}. They
estimate the SN brightness by doing PSF fitting photometry on images where the background light
of the galaxy had been removed using very late, high quality images.
Finally, \citet{fetal00} analyze very late HST images of ESO184-G82 taken through non standard passbands, and provide
estimates of the SN brightness more than 750 days after the GRB.

The main purpose of this paper is to present an independent set of photometric
observations done at Cerro Tololo Inter-American Observatory between June and
October of 1998, from $\sim$40 up to 160 days after the GRB.
The CTIO observations overlap most of the data sets mentioned above, and
our $U$ band observations reach about 30 days later than those published so far.

We purposely calibrated our photometry with the same set of local standards
used by \citet{getal98} for the early light curve, and \citet{soetal02} for
the very late light curve.
Therefore, if $S$-corrections \citep{petal04} between the different realizations of the
passbands are not large, these data sets could be straightforwardly merged.
This provides a secondary goal for this paper.
We merge our data with the earlier set of \citet{getal98} and later set of
\citet{soetal02} providing an ensemble of 300 $UBV(RI)_C$ observations, spanning from about
15 days before up to more than 500 days after maximum light, consistently calibrated with the
sequence of local standards as defined by the CTIO realization of the standard photometric system.
Finally having this photometry available, it is possible to obtain
multicolor, unreddened, rest frame light curves, as well as quasi-bolometric light curves.
We do so providing fits to the photometry that extend more than
$\sim$80 days after the GRB with $UBV(RI)_C$ coverage, and close to 500 days
after the GRB with only $BV(RI)_C$.

In \S \ref{se:observations} we present our observations, in \S
\ref{se:lcurves} we describe the comparison and merging of our data with those already published,
and the computing of $K$ and extinction corrections to obtain the rest frame unreddened light
curve of SN~1998bw, which we present interpolated with low degree piecewise splines.
Finally, in \S \ref{se:conclusions} we summarize our work.

\section{Observations} \label{se:observations}

The CTIO observations started $\sim$40 days after the GRB (about a month after maximum light).
The 0.9m telescope with a direct CCD camera attached was used.
The detector was a TEK~2048 CCD, with a pixel size of 24 $\mu$, providing
a scale of 0.396 arc-seconds per pixel.
The passbands routinely used were the standard $UBV(RI)_C$ of the Johnson, Kron--Cousins photometric
system \citep{kronsmith51, johnson55, cousins76}.
$BV(RI)_C$ images were taken on 15 nights while $U$ images were taken on 12.
Five of those nights appeared to be photometric, and extensive sets of standards from the lists of
\citet{landolt92} were observed as well, to fit color terms for the instrument and extinction
coefficients for the nights.

\begin{figure}[t]
\plotone{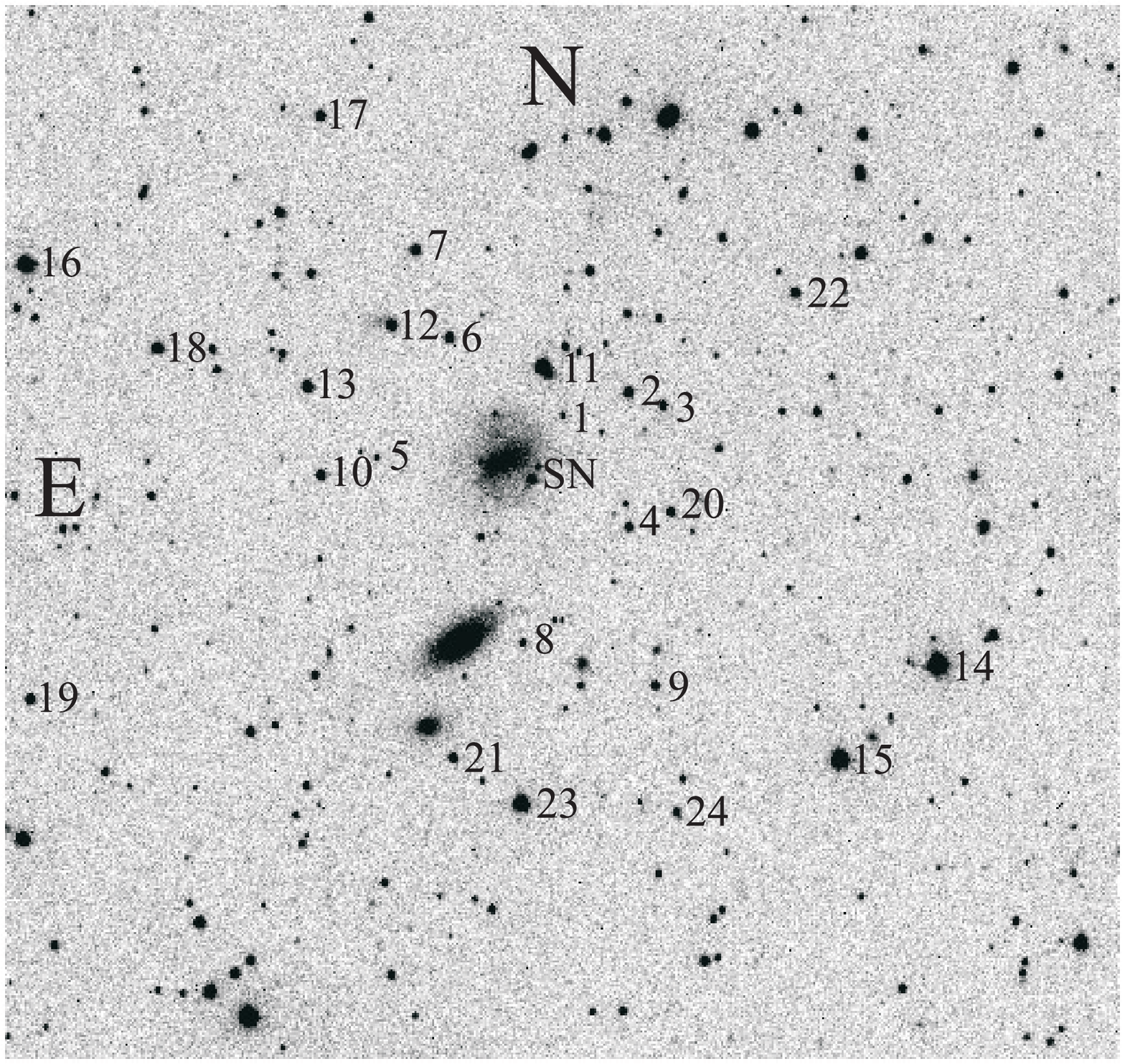}
\caption{Local Standard Sequence for SN~1998bw. The image was taken with CTIO 0.9m telescope on
October 2, 1998, using the $I_{rm C}$ passband. Exposure time was 600 seconds.
North is up and East  is left. The width of the field in the East-West direction
is approximately 6.6 arc-minutes.
Stars number 1 to 15 are the same local sequence of \citet{getal98}.
\label{fi:sequence}}
\end{figure}

Reduction of the images was done in the usual manner within the {\em IRAF}\footnote{The Image Reduction and
Analysis Facility is developed and maintained by NOAO, under
contract with the National Science Foundation.} environment.
Briefly, images were trimmed, bias-corrected, and flat-fielded.
A sequence of isolated stars with good signal to noise ratio was located and used to build a
variable PSF for each image using the package DAOPhot \citep{stetson87}.
The PSF is later fitted to all the stars of the local photometric sequence indicated in Figure
\ref{fi:sequence}, and the SN, to estimate their instrumental magnitudes.
Using the instrumental magnitudes, the
$UBV(RI)_C$ magnitudes for the local sequence and the color terms for the instrument that had been
measured in the photometric nights, we transformed the instrumental PSF magnitudes of the SN
into calibrated magnitudes.

\begin{figure}[t]
\includegraphics[angle=-90,scale=.70]{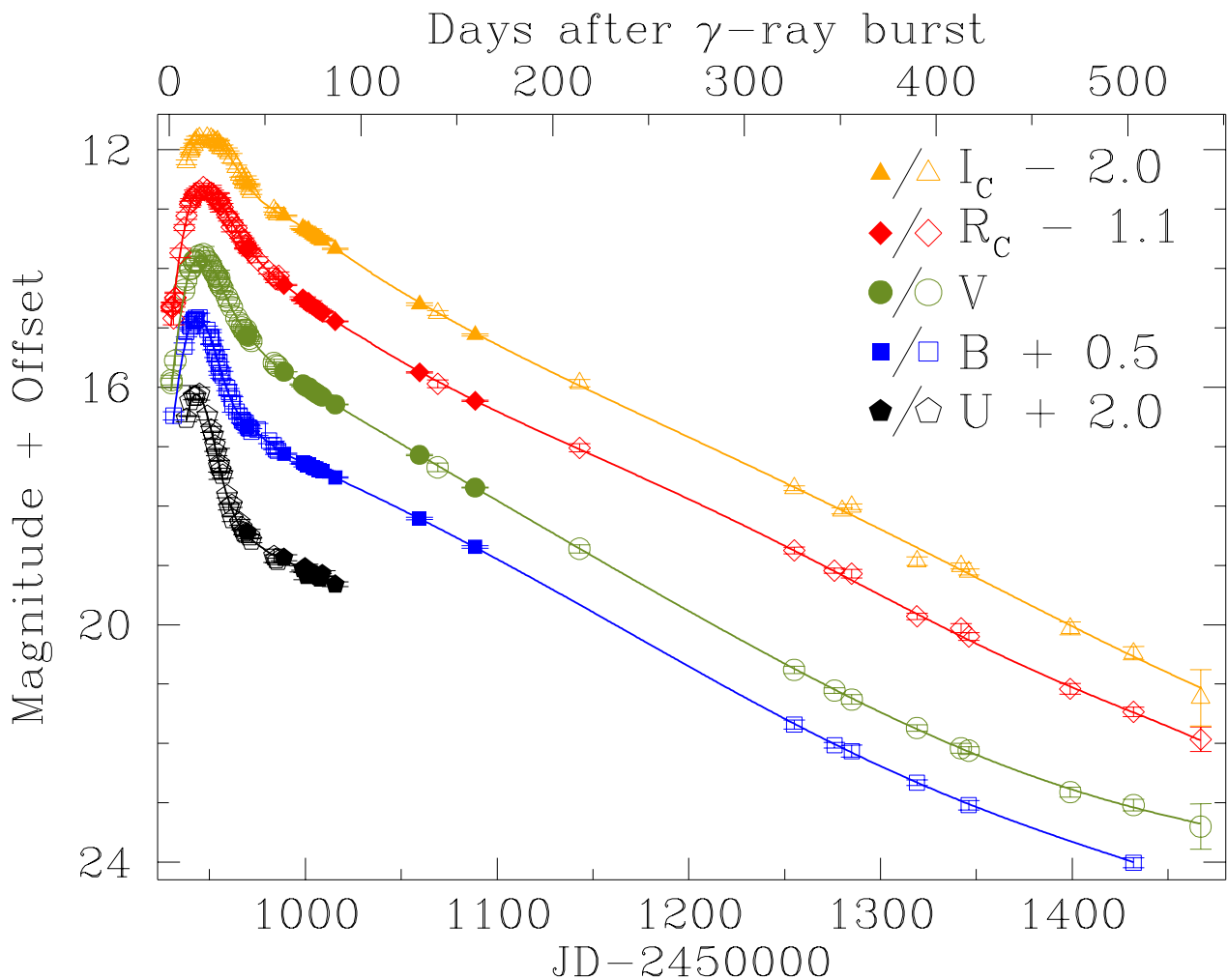}
\caption{Complete light curves of SN~1998bw. Open symbols display the observations of \citet{getal98} at early time and \citet{soetal02} at
late times. Solid symbols bridging the two sets
are the CTIO observations presented here. All the observations were transformed to the photometric system
defined by the local standards in Table \ref{ta:sequence}. The solid lines are the fits used to build the intrinsic light curves given in Table \ref{ta:restframephot}.
\label{fi:phot}}
\end{figure}

After reduction, two of the five supposedly photometric nights gave larger than expected residuals for the photometric fits.
Then, they were used in the fitting of color terms, but not in the absolute calibration of the sequence of
local standards.
The magnitudes of the local sequence of standards is given in Table~\ref{ta:sequence}.
The photometry of SN~1998bw is given in Table~\ref{ta:ctiophot}, under reference code (2), and plotted in Figure \ref{fi:phot}.

\section{Light Curves} \label{se:lcurves}

\subsection{Merged Photometry}

The photometry presented above can be joined with data published elsewhere to build a well sampled and
extended multicolor light curve of SN~1998bw.
After comparing our raw observations with those of \citet{getal98,mands99} and \citet{soetal02},
we decided to join our data with those of \citet{getal98} and \citet{soetal02}.
At early times, the data of \citet{getal98} is unique, and must be considered.
At late times, the data of \citet{soetal02} include a reanalysis of the images used by \citet{petal01}, and
corresponds to PSF photometry on images with the background subtracted.
An additional consideration is that \citet{getal98} and \citet{soetal02} calibrate the photometry using a common
set of 15 local standards while \citet{mands99} use a different local sequence.

After a careful comparison,
we found phase dependent systematic differences between our photometry and that of \citet{mands99}.
They are probably the result of slightly different realizations of the photometric passbands combined with
the non stellar character of the spectrum of SN~1998bw, the different set of local standards used, and the
different technique applied to estimate the SN brightness (aperture versus PSF photometry).
The effects of bandpass differences causing large systematic errors in supernova photometry has
been known for quite a while \citep{setal88}.
Since our earlier data are coincident in time with those of \citet{mands99}, and at late times our data merges
already with those of \citet{soetal02} we decided not to use the observations of the former in our combined
light curve.

\begin{figure}[t]
\includegraphics[angle=0,scale=0.7]{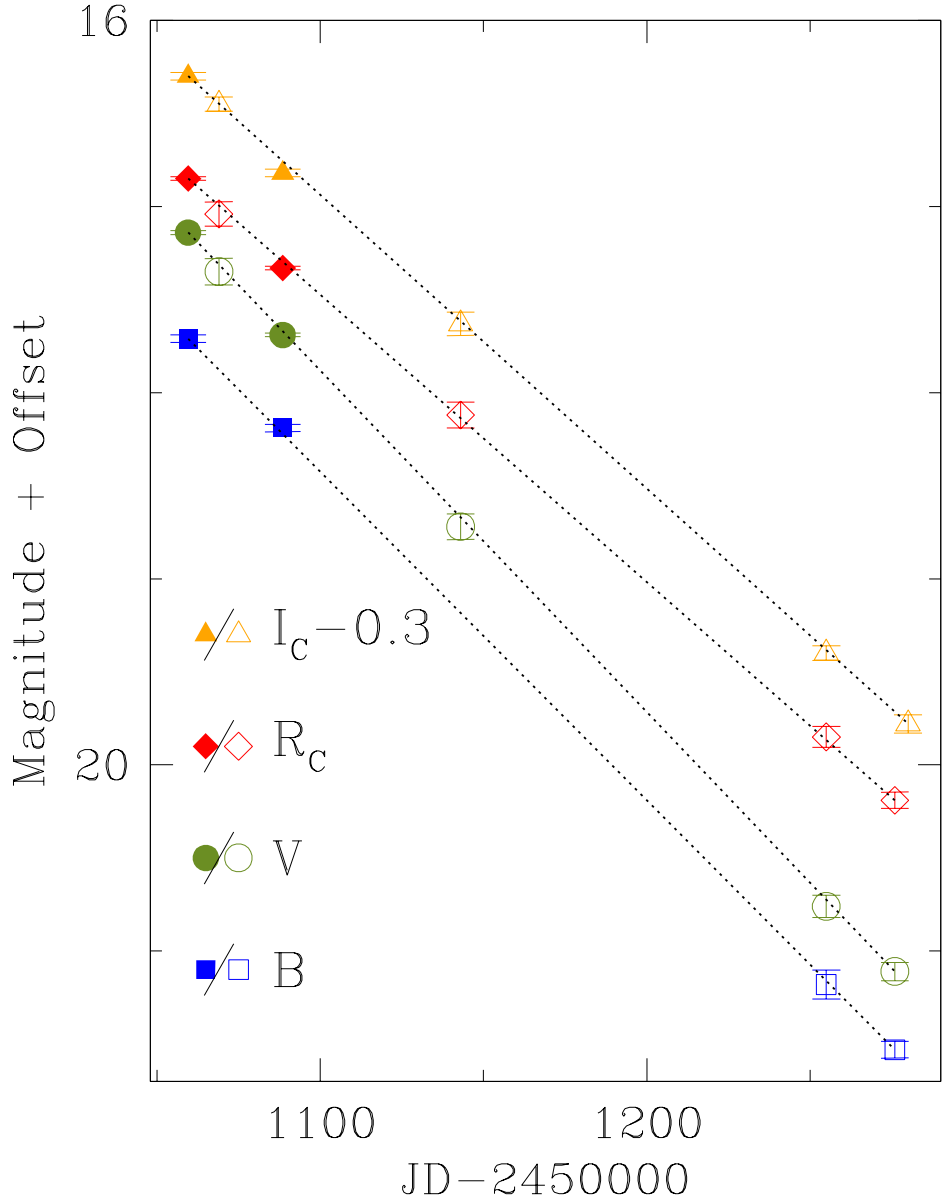}
\caption{Expanded view of the region of overlap between the CTIO observations and those of \citet{soetal02}. Solid symbols
correspond to the last epochs of CTIO photometry (Table \ref{ta:ctiophot} under reference number 2), open symbols to the
first epochs observed by \citet{soetal02}. For each passband, dotted straight lines connect the first and last observation plotted.
\label{fi:join}}
\end{figure}

We also found small systematic differences between our photometry and that of \citet{getal98} and \citet{soetal02} as well.
Within the uncertainties, however, the differences were constant with SN phase.
We took this as an indication of negligible $S$-corrections between our photometry and those of either \citet{getal98} or \citet{soetal02}.
Since the three studies had in common a sequence of 15 local standards, it was possible to fit zero points and relative color terms
between the two sets, and transform all the observations to the system defined by the magnitudes of the local standard set
given in Table~\ref{ta:sequence}.
The transformed magnitudes merge very well.
The transformed photometry from different sources is consistent within the
uncertainties.
It is particularly reassuring that they merge very well at our latest times
in the passbands $R_C$ and $I_C$, which are those where the effect of
the $S$-corrections are expected to be strongest (see Figure \ref{fi:join}).
Since \citet{soetal02}, who obtained the very late time photometry shown in the figure, did correct the observations for background light
contamination and we did not, this
means in addition, that as late as $\sim$160 days after the GRB, contamination by background light was not critically important, and that
our PSF fitted photometry is not biased.

The transformed photometry is given in Table \ref{ta:ctiophot} under reference codes (1) and (3), and plotted in Figure \ref{fi:phot}
together with the new data presented here.
In addition to providing a consistently calibrated bridge between the photometry of \citet{getal98} and that of \citet{soetal02} the CTIO
observations define the slope of the early exponential tail in $U$, which was not known.

\subsection{The Unreddened Rest Frame Light curves}

To compute the intrinsic light curve of SN~1998bw, it is necessary to correct the photometry for the effect of extinction by foreground material
and apply $K$-corrections to transform it to the rest frame of ESO184-G82.
It is always a challenge to compute the extinction towards an object of exotic type, like SN~1998bw, because the intrinsic colors are not known.
All the indications, however, suggest that the amount of foreground matter between us and the SN is small.
The Galactic dust towards ESO184-G82 contributes $A_V \sim 0.2$, according to \citet{sfandd98}, up from the earlier value of $A_V \sim 0.05$
given by \citet{bandh82}.
\citet{petal01} obtained high resolution spectra of SN~1998bw near maximum light. They did not find Na~D absorption lines neither at the redshift of
ESO184-G82 nor zero, and set upper limits to the extinction consistent even with the higher estimate of \citet{sfandd98}.

The peculiarities of the spectrum of SN~1998bw and the fact that it evolves in time prompted us to compute a time dependent extinction in each
passbands using the excellent series of spectra of \citet{petal01}, which is publicly available.
We took the transmittance of the $BV(RI)_C$ passbands given by \citet{bessell90},
transformed them from energy-based to photon-based units, convolved them with
a typical CCD quantum efficiency, the spectral response of two aluminum coated surfaces, and a typical
response of the sky telluric absorption bands, and used this combined sensitivity curves to compute the difference in magnitudes
between the unreddened and reddened spectra of SN~1998bw.
The differences between the extinction computed in this way and a constant are small but they vary with time, allowing for small systematic
changes in the trends of the light curves.

The extinction law towards $\gamma$-ray bursts has deserved some attention \citep[see ][ and references therein]{landl10}. It has been found that, in
some cases, the extinction estimated from the $\gamma$-ray burst spectrum is different from that of the Milky Way or Magellanic Clouds.
When there are differences, however, they are typically
marked in the UV, at wavelengths below $\sim$2500 \AA, and modest at longer wavelengths.
This, together with the low estimate of the extinction towards SN~1998bw, make it
reasonable to assume a typical Milky Way extinction law to unredden the lights curves presented here.
We take as a model the extinction curve of \citet{ccandm89} with $R_V = 3.1$.

Similar considerations apply to the $K$-corrections, which, at the redshift of ESO184-G82 \citep[$z =$ 0.0087,][]{fetal06}, are expected to be small,
but still variable in time.
We used the passbands of \citet{bessell90}, prepared as stated above and the spectra of \citet{petal01} to compute the
time dependent $K$-corrections of SN~1998bw.

Using the computed reddening and $K$ corrections unreddened and transformed to the rest frame the photometry of SN~1998bw
given in Table~\ref{ta:ctiophot}.
The results, fitted with
low-degree piecewise polinomial and/or spline fits to provide daily sampling, are presented in Table~\ref{ta:restframephot}.
The uncertainties given in the table are the result of combining in quadrature the uncertainties of photometry, reddening and $K$-corrections.

Having the multicolor rest frame photometry it is possible to 
convert the observed single-bandpass light curves into
monochromatic fluxes at the effective wavelengths of the passbands,
integrate them over frequency, and obtain quasi-bolometric light curves
that include the energy output in all the observed bands
\citep{sandb90}.
We used the zero-points of \citet{bessell00} and integrated the monochromatic fluxes using a trapezoid rule.
We did not extrapolate the flux beyond the limits of the $U$ (or $B$) and $I_C$ passbands.
The distance to ESO~184-G082 was taken to be $37.3 \pm 2.6$ Mpc, as computed by the NED database\footnote{http://nedwww.ipac.caltech.edu/}
from the redshift measured by \citet{fetal06},
a Hubble constant of 74.2 km s$^{-1}$ Mpc$^{-1}$ \citep{retal09}, and a the model of the local velocity field given by \citet{metal00}.
The $UBV(RI)_C$ and $BV(RI)_C$ luminosities have been included in Table~\ref{ta:restframephot}.

\section{Conclusions} \label{se:conclusions}

We present 72 new photometric observations of SN~1998bw in the $UBV(RI)_C$ bands, spanning from $\sim$40 up to $\sim$60 days after the GRB.
Our $U$ data extends by about 30 days the previously known coverage and sample the early exponential tail.

We collect previously published photometry and transform it to our realization of the standard system using relative zero points and color
terms measured from the common local standard sequence.
We show that data from different sources merges smoothly, providing a homogeneous set of 300 observations covering about 500 days since explosion.

Finally, we use the extensive series of spectra in public domain to compute time dependent reddening and $K$ corrections and compute
the unreddened rest frame multicolor and quasi-bolometric light curves. We fit them with low degree piecewise splines to provide daily sampling
and give the results as a table.

\acknowledgments

We thank the staffs of the CTIO observatory for their assistance with the observations and
Arlo Landolt for sending us his tables of photometric standards prior to publication.
AC acknowledges the support from grants P06-045-F (ICM, MIDEPLAN, Chile), Basal CATA PFB 06/09, and FONDAP No. 15010003.
This research has made use of the SUSPECT on-line database of SN spectra and of the NASA/IPAC
Extragalactic Database (NED) which is operated by the Jet Propulsion Laboratory, California Institute
of Technology, under contract with the National Aeronautics and Space Administration.

\newpage

\clearpage

\clearpage

\clearpage

\begin{deluxetable}{lccccc}
\tablewidth{0pt}
\tablecaption{Local Standard Sequence for SN~1998bw\label{ta:sequence}}
\tablehead{
\colhead{ID\tablenotemark{a}} &
\colhead{$V$}  &
\colhead{$B-V$}  &
\colhead{$U-B$}  &
\colhead{$V-Rc$} &
\colhead{$V-Ic$} 
}
\startdata
  1 & 18.503  $\pm$   0.032 &  0.422  $\pm$   0.053 & -0.208  $\pm$   0.053 &  0.293  $\pm$   0.050 &  0.576  $\pm$   0.043 \\
  2 & 17.696  $\pm$   0.006 &  1.161  $\pm$   0.044 &  0.964  $\pm$   0.179 &  0.717  $\pm$   0.010 &  1.309  $\pm$   0.007 \\
  3 & 18.072  $\pm$   0.008 &  0.864  $\pm$   0.031 &  0.570  $\pm$   0.041 &  0.495  $\pm$   0.023 &  0.972  $\pm$   0.012 \\
  4 & 17.792  $\pm$   0.022 &  0.696  $\pm$   0.024 &  0.085  $\pm$   0.039 &  0.399  $\pm$   0.023 &  0.780  $\pm$   0.025 \\
  5 & 19.403  $\pm$   0.021 &  0.772  $\pm$   0.030 &         \nodata       &  0.435  $\pm$   0.029 &  0.855  $\pm$   0.041 \\
  6 & 17.215  $\pm$   0.014 &  0.641  $\pm$   0.016 &  0.022  $\pm$   0.013 &  0.397  $\pm$   0.016 &  0.761  $\pm$   0.016 \\
  7 & 17.489  $\pm$   0.021 &  0.831  $\pm$   0.022 &  0.315  $\pm$   0.016 &  0.524  $\pm$   0.022 &  1.024  $\pm$   0.021 \\
  8 & 18.572  $\pm$   0.018 &  0.622  $\pm$   0.021 & -0.025  $\pm$   0.034 &  0.394  $\pm$   0.051 &  0.753  $\pm$   0.029 \\
\enddata
\tablenotetext{a}{Stars number 1 to 15 are the same as those of \citet{getal98}.}
\tablenotetext{b}{Star number 25 falls outside the finding chart of Figure \ref{fi:sequence}.
It is located 23.1 arc-seconds North and 20.6 arc-seconds East from star number 19.}
\end{deluxetable}

\clearpage

\begin{deluxetable}{rcccccc}
\tablewidth{0pt}
\tablecaption{Photometry of SN~1998bw transformed to CTIO System\label{ta:ctiophot}}
\tablehead{
\colhead{JD\tablenotemark{a}} &
\colhead{$U$}  &
\colhead{$B$}  &
\colhead{$V$}  &
\colhead{$Rc$} &
\colhead{$Ic$} &
\colhead{Ref.\tablenotemark{b}}
}
\startdata
 930.13 &               \nodata &               \nodata &   15.88 $\pm$    0.08 &   15.74 $\pm$    0.07 &               \nodata & (1) \\ 
 930.25 &               \nodata &               \nodata &               \nodata &   15.77 $\pm$    0.06 &               \nodata & (1) \\ 
 930.31 &               \nodata &               \nodata &   15.93 $\pm$    0.08 &   15.74 $\pm$    0.07 &               \nodata & (1) \\ 
 931.30 &               \nodata &   15.98 $\pm$    0.07 &               \nodata &   15.94 $\pm$    0.11 &               \nodata & (1) \\ 
 931.92 &               \nodata &               \nodata &               \nodata &   15.69 $\pm$    0.06 &               \nodata & (1) \\ 
 932.19 &               \nodata &               \nodata &   15.55 $\pm$    0.08 &   15.59 $\pm$    0.07 &               \nodata & (1) \\ 
 932.22 &               \nodata &               \nodata &               \nodata &   15.59 $\pm$    0.09 &               \nodata & (1) \\ 
 934.83 &               \nodata &               \nodata &               \nodata &   14.84 $\pm$    0.08 &               \nodata & (1) \\
\enddata
\tablenotetext{a}{Mean Julian Date minus 2450000.}
\tablenotetext{b}{References for photometry are: (1) \citet{getal98}, (2) This work, (3) \citet{soetal02}.}
\end{deluxetable}

\begin{deluxetable}{rrccccccc}
\rotate
\tablewidth{0pt}
\tablecaption{Intrinsic Light Curve of SN~1998bw \label{ta:restframephot}}
\tablehead{
\colhead{JD\tablenotemark{a}} &
\colhead{Phase\tablenotemark{b}} &
\colhead{$U_0$}  &
\colhead{$B_0$}  &
\colhead{$V_0$}  &
\colhead{$Rc_0$} &
\colhead{$Ic_0$} &
\colhead{$F_{UBV(RI)_C}$\tablenotemark{c}} &
\colhead{$F_{BV(RI)_C}$\tablenotemark{c}}
}
\startdata
938 & 8.5 & 14.20 $\pm$ 0.04 & 14.36 $\pm$ 0.12 & 13.98 $\pm$ 0.08 & 14.03 $\pm$ 0.08 & 14.05 $\pm$ 0.01 & 520.973 $\pm$ 39.134 & 403.065 $\pm$ 29.406 \\
939 & 9.5 & 14.04 $\pm$ 0.04 & 14.25 $\pm$ 0.06 & 13.86 $\pm$ 0.06 & 13.91 $\pm$ 0.06 & 13.95 $\pm$ 0.01 & 581.517 $\pm$ 27.301 & 448.978 $\pm$ 21.199 \\
940 & 10.5 & 13.91 $\pm$ 0.06 & 14.16 $\pm$ 0.01 & 13.76 $\pm$ 0.03 & 13.80 $\pm$ 0.05 & 13.87 $\pm$ 0.01 & 636.446 $\pm$ 17.171 & 491.197 $\pm$ 13.565 \\
941 & 11.5 & 13.83 $\pm$ 0.02 & 14.10 $\pm$ 0.01 & 13.67 $\pm$ 0.01 & 13.72 $\pm$ 0.02 & 13.80 $\pm$ 0.01 & 683.884 $\pm$ 6.995 & 528.501 $\pm$ 5.135 \\
942 & 12.5 & 13.78 $\pm$ 0.03 & 14.05 $\pm$ 0.01 & 13.61 $\pm$ 0.01 & 13.64 $\pm$ 0.01 & 13.75 $\pm$ 0.01 & 722.256 $\pm$ 6.137 & 559.822 $\pm$ 4.203 \\
943 & 13.5 & 13.75 $\pm$ 0.04 & 14.02 $\pm$ 0.01 & 13.56 $\pm$ 0.03 & 13.59 $\pm$ 0.02 & 13.70 $\pm$ 0.01 & 749.795 $\pm$ 12.392 & 583.606 $\pm$ 9.806 \\
944 & 14.5 & 13.76 $\pm$ 0.01 & 14.02 $\pm$ 0.01 & 13.53 $\pm$ 0.03 & 13.55 $\pm$ 0.02 & 13.67 $\pm$ 0.02 & 766.461 $\pm$ 13.486 & 600.031 $\pm$ 11.953 \\
945 & 15.5 & 13.79 $\pm$ 0.03 & 14.03 $\pm$ 0.01 & 13.51 $\pm$ 0.01 & 13.52 $\pm$ 0.02 & 13.65 $\pm$ 0.01 & 772.324 $\pm$ 8.626 & 608.760 $\pm$ 6.209 \\
\enddata
\tablenotetext{a}{Mean Julian Date minus 2450000.}
\tablenotetext{b}{Rest frame days after $\gamma$-ray burst.}
\tablenotetext{c}{Units are $10^{40}$ erg cm$^{-2}$ s$^{-1}$. Uncertainty does not include the $\sim$7\% uncertainty in the distance.}
\end{deluxetable}
\end{document}